%% file: main.tex
\documentclass{article}

\usepackage{arxiv}

\usepackage[utf8]{inputenc} % allow utf-8 input
\usepackage[T1]{fontenc}    % use 8-bit T1 fonts
\usepackage{hyperref}       % hyperlinks
\usepackage{url}            % simple URL typesetting
\usepackage{booktabs}       % professional-quality tables
\usepackage{amsfonts}       % blackboard math symbols
\usepackage{nicefrac}       % compact symbols for 1/2, etc.
\usepackage{microtype}      % microtypography
\usepackage{lipsum}         % Can be removed after putting your text content
\usepackage{graphicx}
\usepackage{doi}

\usepackage{latexsym}
\usepackage{graphicx}
\usepackage{multicol,multirow}
\usepackage{amsmath,amssymb,amsfonts}
\usepackage{mathtools}
\usepackage{mathrsfs}
\usepackage{amsthm}
\usepackage{nicefrac}
\usepackage{rotating}
\usepackage{appendix}
\usepackage[most]{tcolorbox}
\usepackage[sort, numbers]{natbib}
\usepackage{ifpdf}
\usepackage[T1]{fontenc}
\usepackage{times}
\usepackage{hyperref}
\usepackage{xcolor}
\usepackage{multirow}
\usepackage{csquotes}
\usepackage{newtxmath}
\usepackage{textcomp}%
\usepackage{xcolor}%
\usepackage{hyperref}
\usepackage{lipsum}
\usepackage{makecell}
\usepackage{float}
\input{lettres_maths}

\title{AODisaggregation: toward global aerosol vertical profiles}

\date{}

\author{\href{https://orcid.org/0000-0001-8068-8527}{\hspace{1mm}Shahine Bouabid}\thanks{Correspondence to \texttt{shahine.bouabid@stats.ox.ac.uk}} \\
	Department of Statistics\\
	University of Oxford \\
	Oxford, UK \\
	\And
	\href{https://orcid.org/0000-0002-5312-4950}{\hspace{1mm}Duncan Watson-Parris} \\
	Atmospheric, Oceanic and Planetary Physics\\
	Department of Physics \\
	University of Oxford\\
	Oxford, UK \\
	\And
	{\hspace{1mm}Sofija Stefanović} \\
	Department of Earth Sciences \\
	University of Cambridge\\
	Cambridge, UK \\
	\And
	\href{https://orcid.org/0000-0003-3873-9970}{\hspace{1mm}Athanasios Nenes} \\
	Laboratory of Atmospheric Processes and their Impact\\
	École Polytechnique Fédérale de Lausanne\\
	Lausanne, Switzerland \\
	\And
	\href{https://orcid.org/0000-0001-5547-9213}{\hspace{1mm}Dino Sejdinovic} \\
	Department of Statistics\\
	University of Oxford \\
	Oxford, UK \\
}

\renewcommand{\shorttitle}{AODisaggregation: toward global aerosol vertical profiles}

\hypersetup{
pdftitle={AODisaggregation: toward global aerosol vertical profiles},
pdfsubject={q-bio.NC, q-bio.QM},
pdfauthor={Shahine Bouabid, Duncan Watson-Parris, Sofija Stefanović, Athanasios Nenes, Dino Sejdinovic},
pdfkeywords={AOD, aerosol extinction profiles, vertical disaggregation, Gaussian processes, ECHAM-HAM},
}

\begin{document}
\maketitle

\vspace*{-2em}
\begin{abstract}
	Aerosol-cloud interactions constitute the largest source of uncertainty in assessments of the anthropogenic climate change. This uncertainty arises in part from the difficulty in measuring the vertical distributions of aerosols, and only sporadic vertically resolved observations are available. We often have to settle for less informative \emph{vertically aggregated} proxies such as aerosol optical depth (AOD). In this work, we develop a framework for the vertical disaggregation of AOD into extinction profiles, i.e. the measure of light extinction throughout an atmospheric column, using readily available vertically resolved meteorological predictors such as temperature, pressure or relative humidity. Using Bayesian nonparametric modelling, we devise a simple Gaussian process prior over aerosol vertical profiles and update it with AOD observations to infer a distribution over vertical extinction profiles. To validate our approach, we use ECHAM-HAM aerosol-climate model data which offers self-consistent simulations of meteorological covariates, AOD and extinction profiles. Our results show that, while very simple, our model is able to reconstruct realistic extinction profiles with well-calibrated uncertainty, outperforming by an order of magnitude the idealized baseline which is typically used in satellite AOD retrieval algorithms. In particular, the model demonstrates a faithful reconstruction of extinction patterns arising from aerosol water uptake in the boundary layer. Observations however suggest that other extinction patterns, due to aerosol mass concentration, particle size and radiative properties, might be more challenging to capture and require additional vertically resolved predictors.
\end{abstract}

\vspace*{1em}
\begin{impact}
	Aerosol-cloud interactions (ACIs) represent the largest uncertainty in assessments of global warming, with uncertainty bounds that could offset global warming or double its effects. This uncertainty arises in part from the inability to observe aerosol amounts at an appropriate resolution. Instead aerosol optical depth (AOD) --- observed globally through satellite products --- is commonly used as a two-dimensional proxy for aerosols. Yet, obtaining a global estimate of aerosol vertical distribution in the atmosphere would be more informative and help better constrain ACIs uncertainty. Here, we show that using AOD and readily available vertically resolved meteorological predictors (temperature, pressure, relative humidity, updraft), a simple probabilistic model can already yield realistic vertical extinction profiles together with appropriate uncertainty quantification. This highlights that such simple modelling can benefit satellite products, leading to more accurate priors over aerosol vertical profiles.
\end{impact}

\section{Introduction}\label{section:introduction}

Aerosols are microscopic particles ($<5\,\mu m$) suspended in the atmosphere. They can come from natural sources (e.g.\ dust, sea salt) or be emitted by human activity (e.g.\ black carbon).

They influence the Earth's energy budget with a negative radiative forcing that counteracts the global warming from anthropogenic greenhouse gases emissions. A fraction of this negative forcing stems from aerosols' direct scattering of incoming solar radiation~\cite{mccormick1967climate}: this is the \emph{direct effect}. A larger fraction of this forcing is due to their modulation of radiative properties of clouds: this is the \emph{indirect effect}. By acting as cloud condensation nuclei (CCN), additional aerosols can drive up the cloud droplet number while driving down the mean cloud droplet size. The resulting clouds are brighter, larger and last longer~\cite{twomey1977influence, albrecht1989aerosols}. They hence reflect more solar radiation and cool the Earth.

Unlike the direct effect which can, in principle, be well constrained~\cite{watson2020constraining}, the magnitude of the forcing induced by the indirect effect is difficult to estimate. There are two reasons for this: (1) the physical processes underpinning aerosol-cloud interactions (ACIs) are not yet fully understood, which hinders the estimation of present day forcing; (2) the present day forcing must be compared to the forcing at pre-industrial state, which is also particularly challenging to quantify~\cite{carslaw2013large}. In fact, observational and model-based studies of ACIs still disagree on the magnitude of this forcing. As a result, ACIs contribute the largest uncertainty in present day global warming~\cite{ipcc2021}.

To better estimate present day forcing we require accurate, global measurements of CCN concentrations to assess radiative properties of clouds~\cite{twomey1977influence, albrecht1989aerosols}. Unfortunately, measuring CCN concentrations can only be achieved in-situ, and while field campaigns have already undertaken to collect detailed CCN observations, these measurements are spatio-temporally sparse and provide insufficient constraint on global distribution of aerosols~\cite{andreae2009correlation, spracklen2011global}.

For lack of better observations, the Aerosol Optical Depth (AOD) has been widely adopted as a first order proxy of CCN concentration in an atmospheric column~\cite{nakajima2001possible, andreae2009correlation, clarke2010hemispheric}. The AOD is a measure of the extinction of solar radiations through an atmospheric column. It is denoted by $\tau$ and defined at a given wavelength, time, latitude and longitude by
\begin{equation}\label{eq:aod}
    \tau = \int_0^H b_\text{ext}(h)\d h,
\end{equation}
where $b_\text{ext}$ is the extinction coefficient\footnote{the sum of contribution from particle-light scattering plus absorption of light by particles} and the integral is taken over the height $H$ of an atmospheric column. The AOD is appealing because it is routinely observed on a global scale by satellite products~\cite{remer2005modis} which, as opposed to in situ observations, offer long term global records.

However, the AOD is a column-integrated quantity and does not provide information on the vertical distribution of aerosols. This is limiting as their vertical distribution strongly influences both the magnitude and even the sign of the forcing induced by the indirect effect. For example, both modelling~\cite{stier2016limitations} and observational studies~\cite{painemal2020reducing} find AOD inadequate for assessing ACIs over vast subtropical ocean areas, which play a key role in determining the radiation balance of the Earth. Yet, in both studies, the vertically resolved aerosol extinction coefficient $b_\text{ext}$ shows significantly higher correlation with CCN concentrations. \citet{stier2016limitations} also highlights the importance of determining aerosol vertical distributions to provide stronger constraints on CCN at specific altitudes. In particular, the AOD fails to describe near-surface properties such as the concentration of aerosols in the boundary layer.

In this work, we propose to probe whether AOD observations can be used to constrain a global prior over aerosol vertical distributions. Formally, given an AOD observation $\tau$, we want to reconstruct the corresponding extinction coefficient profile $b_\text{ext}$. This amounts to the task of reconstructing three-dimensional (3D) profiles using height-integrated two-dimensional (2D) observations and quantities that are easier to obtain in 3D, such as temperature and relative humidity.

Motivated by the study of cloud vertical structures, this task has been framed in the past as fully-supervised learning~\cite{leinoen2019reconstruction}, i.e.\ assuming observations of groundtruth vertical profiles were available. Collecting vertically resolved observations of aerosols optical properties is also possible, using lidar-based remote sensing instruments~\cite{winker2013global} or groundbased sun-photometers~\cite{holben1998aeronet}. While valuable, these observations are however limited by their low spatiotemporal coverage and prone to corruption (e.g.\ low signal-to-noise ratio, clear-sky requirement). Compiling high quality observational data of aerosol vertical profiles at large scale is thus challenging, making fully-supervised learning approaches inadequate.

Instead, we propose to draw from spatial disaggregation methodologies that only require observations at the aggregated level. Spatial disaggregation is the task of inferring subgrid details given coarse resolution spatial observations. Postulating an underlying fine grained spatial field that aggregates into coarse observations, this problem can be framed as weakly supervised learning~\cite{zhou2017weakly} with aggregated targets. While existing works~\cite{zhang2020aggregate, law2018variational, yousefi2019multitask, tanaka2019spatially, tanskanen2020nonlinearities} have only considered aggregation processes happening on a 2D field, this rationale can be extended to disaggregate quantities along a third dimension --- height. Since $\tau$ corresponds to the vertical integration of $b_\text{ext}$, we propose to frame the reconstruction of aerosol vertical profiles as the vertical disaggregation of AOD observations.

Using Gaussian processes (GPs)~\cite{rasmussen2005gaussian}, we design a Bayesian model that maps vertically resolved meteorological variables (e.g.\ pressure, relative humidity) to a probabilistic estimate of the extinction coefficient that integrates into the AOD. The model formulation is simple and makes assumptions explicit, hence granting control and interpretability over predictions while offering built-in uncertainty quantification.

In order to be able to fully validate the proposed methodology, we use ECHAM-HAM global aerosol-climate model simulation data~\cite{stier2005aerosol, stier2007aerosol, zhang2012global}. While our prime motivation is to reconstruct $b_\text{ext}$ from satellite observations of AOD, the intricacies of combining measurements from different instruments makes it challenging to validate any proposed methodology. On the other hand, ECHAM-HAM is a self-consistent climate model that offers readily available aerosol vertical profiles, and is better suited for model development. We demonstrate our model is able to reconstruct natural patterns that arise in aerosol vertical distribution, in particular in the boundary layer. We show that very simple and readily available meteorological predictors suffice to obtain a good estimation of the extinction coefficient.

The aims of this study are as follows:
\begin{itemize}
    \item Formulate a probabilistic vertical disaggregation methodology for the task of reconstructing aerosol vertical profiles using readily available vertically resolved covariates.
    \item Validate the proposed methodology using climate model data --- where access to groundtruth extinction coefficient profiles enables quantitative evaluation.
\end{itemize}

Section~\ref{section:model} outlines the design of the vertical disaggregation model and describes the inference procedure as well as the hyperparameter selection. Section~\ref{section:experimental-setup} describes the dataset and experimental setup used to validate the model. Section~\ref{section:results} presents the experimental results and Section~\ref{section:discussion} discusses, while introducing avenues of future research.

\section{Model Design}\label{section:model}
In this section, we first outline the design of the prior distribution we place on the extinction coefficient profile using GPs. We then describe the observation model that connects our prior to the observations of the AOD. Finally, we present how the inference is conducted.

\subsection{Design of the Prior}\label{subsection:prior}

\subsubsection{An idealized vertical prior}
In passive satellite sensors, AOD retrieval algorithms need to assume a form for the vertical profiles. These are commonly idealized, assuming in the simplest case an exponential profile $b_\text{ext}(h)\propto e^{-h/L}$~\cite{li2020impact, wu2017impact}. $L$ is a fixed height scale parameter that is typically taken as the top altitude of the boundary layer ($2\,km$), although in practice different algorithms assume different values.

While idealized, these profiles capture a key element of aerosol vertical distribution: most CCN lies at low altitude in the boundary layer. Hence, we decide to use an idealized exponential profile as a component of the structural prior placed on the extinction profile.

\subsubsection{Weighting the ideal profile}

Clouds' local meteorology influences clouds properties, and hence the sign and magnitude of the effective radiative forcing due to ACIs~\cite{ackerman2004impact, small2009can, chen2014satellite, douglas2020quantifying} --- this source of heterogeneity is at the heart of the uncertainty over how clouds impact climate.

The impact of local meteorology on ACIs can be characterized by the following set of environmental confounders~\cite{kohler1936nucleus, twomey1974}: temperature $T$, pressure $P$, relative humidity RH and vertical draft or updraft $\omega$. Standing as a proxy for aerosols, the AOD is also impacted by its surrounding meteorology~\cite{christensen2017unveiling, jesson2022scalable}. These meteorological variables should thus also modulate the extinction coefficient.

A great advantage of these simple meteorological variable is that they (or their proxies) are readily available on multiple pressure levels in reanalysis data. Reanalysis data, such as ERA5~\cite{munoz2021era5}, combines observations and model simulations through physical laws to provide the most accurate representation of past and present climate and meteorology. Hence, $T$, $P$, RH and $\omega$ can be reliably used as vertically resolved predictors.

Finally, note that $b_\text{ext}$ is also a spatiotemporal field which exhibits smoothness across spatial and temporal dimensions. Such regularity should be included in the modelling.

Let $x|h$ denote a $d$-dimensional vector resulting from the concatenation of spatiotemporal and meteorological variables for a given altitude $h$. For example, one can take
\begin{equation}
    x = (t, \text{lat}, \text{lon}, T, P, \text{RH}, \omega),
\end{equation}
where lat and lon respectively denote latitude and longitude. We will denote $\cX\subseteq\RR^d$ the space in which $x$ takes values.

We propose to model the extinction coefficient $b_\text{ext}$ by weighting an idealized exponential vertical profile with a positive weight function $w : \cX\to (0, +\infty)$. Namely, the chosen prior for the extinction coefficient profile is denoted $\varphi$ and takes the simple form
\begin{equation}
    \varphi(x|h) = w(x|h)e^{-h/L}.
\end{equation}
This weight function is meant to capture finer details of variability in the extinction coefficient profile, putting more mass in regions where meteorological predictors suggest aerosol loading is likely to be higher.

\subsubsection{Probabilistic modelling of the weighting function}
While the extinction coefficient should indeed be modulated by $x|h$, we expect this relationship to be non-trivial and highly non-linear~\cite{jesson2022scalable}. For this reason, we propose to learn the weighting function $w$ using non-linear (e.g.\ kernel-based) statistical machine learning methodologies. Furthermore, we want the prior to reflect our lack of knowledge about the relationship between $x|h$ and $b_\text{ext}(h)$. To account for this epistemic uncertainty, we propose a Bayesian formulation of the weighting function.

GPs~\cite{rasmussen2005gaussian} are a ubiquitous class of expressive Bayesian priors over real-valued functions. They have been widely used in various nonlinear and nonparametric regression problems in geosciences~\cite{campvalls2016survey}. A GP is fully determined by its mean function and its covariance function. The covariance function  --- called kernel --- is typically user-specified as a positive definite bivariate function on the input data.

We place a GP prior over the weight function. To ensure the weights are strictly positive, we further warp the GP with a positive transform $\psi : \RR\to (0,+\infty)$. Formally, let $m:\cX\to\RR$ be a mean function and $k : \cX\times\cX\to\RR$ denote a positive definite kernel, we model the weighting function as
\begin{equation}
    w(x|h) = \psi(f(x|h)) \enspace\text{ where}\enspace f\sim\operatorname{GP}(m,k).
\end{equation}
The Bayesian prior defined by $f$ represents a latent variable that maps through the positive link function $\psi$ onto the weight function. A simple choice for $\psi$ is the exponential function, which we will use in Section~\ref{section:experimental-setup}.

While $\psi\circ f$ describes an expressive probability distribution over nonlinear positive functions, it remains interpretable. The choice of the GP kernel $k$ specifies how covariance is encoded, i.e.\ $\operatorname{Cov}\big(f(x), f(x')\big) = k(x, x')$. It also allows to restrict the functional class the GP belongs to. For example, the Matérn class of covariance functions offers control over the functional smoothness of the GP~\cite{rasmussen2005gaussian, stein1999interpolation}.

\subsection{Design of the Observation Model}\label{subsection:observation-model}
We now connect the extinction coefficient prior $\varphi(x|h)$ constructed in Section~\ref{subsection:prior} to the observations of the AOD. For simplicity, we will consider a single air column of height $H$ with observed AOD $\tau$. 

Ideally, we would want to have exactly $\tau = \int_0^H\varphi(x|h)\d h$. Unfortunately, this is unrealistic as AOD observations are prone to distortion. This distortion stems from observational noise~\cite{remer2005modis} as well as from assumptions made in AOD retrieval algorithms~\cite{mielonen2011evaluating, wu2016sensitivity}. It is thus important to account for the measurement error with a probabilistic observation model for $\tau$.

\subsubsection{A log-normal observation model for the AOD}
\begin{figure}
    \centering
    \includegraphics[width=\linewidth]{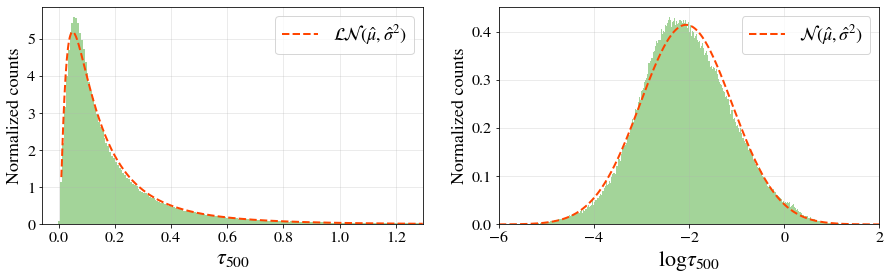}
    \caption{\textbf{Left:} Log-normal density (red) fitted with maximum likelihood estimates to the empirical distribution of AERONET AOD at 500nm ($\tau_{500}$) from 1315 stations between 1993 and 2021 (green). \textbf{Right:} logspace plot of the left panel. It demonstrates a sound fit for the normal distribution in the logspace, albeit with a slight right-skew; $\hat\mu = -2.06$, $\hat\sigma = 0.96$}
    \label{fig:lognormal-fit}
\end{figure}

Since the AOD is strictly positive and highly-skewed toward small values ($\approx\,$0.14), the log-normal distribution has been reported to provide a good fit, e.g. for studies focusing on locations in North America and Europe~\cite{oneil2000lognormal}. The log-normal distribution is a right-skewed continuous probability distribution with support over $(0, +\infty)$. It is specified by a location parameter $\mu\in\RR$ and a scale parameter $\sigma > 0$. In particular, if $Z\sim\cN(0, 1)$ is a standard normal random variable, then $e^{\mu + \sigma Z}$ is a log-normal random variable with location $\mu$ and scale $\sigma$. Using AERONET sun-photometers AOD measurements from 1315 stations between 1993 and 2021~\cite{holben1998aeronet}, we assess the soundness of this observation model. Figure~\ref{fig:lognormal-fit} illustrates that a log-normal density can indeed be appropriately fitted to the empirical distribution of AERONET observations.

Hence, we assume that AOD observations follow a log-normal observation model, which we denote $\cL\!\cN(\mu, \sigma)$. To relate $\tau$ to the prior $\varphi(x|h)$, we use a mean reparametrization of the log-normal distribution. Namely, we know that the expectation of a log-normal random variable is given by $e^{\mu + \sigma^2/2}$. Let $\eta = \EE[\tau] > 0$ be the log-normal mean, then we use the mean-reparametrized observation model given by
\begin{align}
    \tau|\eta & \sim \cL\!\cN\left(\log\eta - \frac{\sigma^2}{2}, \sigma\right) \\
    \eta & = \int_0^H \varphi(x|h)\d h.
\end{align}
When working with multiple AOD observations $\tau_1, \ldots, \tau_n$, the scale parameter $\sigma$ will be assumed shared among atmospheric columns. The mean parameter $\eta$ (or location parameter $\mu$ for the canonical parametrization) will however be column-specific.

\subsubsection{An observation model for $b_\text{ext}$}

While we do not observe $b_\text{ext}(h)$ --- and do not need it for tuning or inference --- it is still useful to specify an observation model for the extinction coefficient. Indeed, at inference stage, $\varphi(x|h)$ will only ever stand as a latent representation of actual observations of the extinction coefficient that integrates in expectation to the AOD. As such, it might fail to capture observational noise which will inevitably hinder variance calibration for the predicted extinction coefficient profile distribution. Using an observation model over $b_\text{ext}$, we can introduce an additional degree of freedom that will help calibrate the variance against a small amount of extinction coefficient observations.

For this reason, we also assume a log-normal observation model for the extinction coefficient\footnote{with this observation model, we do not satisfy exactly $\tau = \int_0^H b_\text{ext}(h)\d h$ as the integral of log-normal random variables is not log-normal. We however choose to keep and overload notations for the sake of the presentation.}
\begin{equation}\label{eq:bext-observation-model}
    b_\text{ext}(h)|\varphi(x|h)\sim \cL\!\cN\left(\log\varphi(x|h) - \frac{\sigma_\text{ext}^2}{2}, \sigma_\text{ext}\right).
\end{equation}
This observation model conserves the mean\footnote{$\EE[b_\text{ext}(h)] = \EE\left[\EE[b_\text{ext}(h)|\varphi(x|h)]\right] = \EE\left[\exp\left(\log\varphi(x|h) - \frac{\sigma_\text{ext}^2}{2} + \frac{\sigma_\text{ext}^2}{2}\right)\right] = \EE[\varphi(x|h)]$} and simply includes observational noise through the scale parameter $\sigma_\text{ext}$. This parameter is calibrated following the procedure described in Section~\ref{subsubsection:sigmaext-calibration}. This will be of particular relevance to validate the model against groundtruth extinction coefficient profiles.

\subsection{Tuning and Inference}

\subsubsection{Finite-sample problem formulation}

We now make things concrete and assume we observe the AOD for $n$ independent columns, which we stack into the vector $\btau = \begin{bmatrix}\tau_1 & \hdots & \tau_n\end{bmatrix}^\top\in\RR^n$. For the $i$\textsuperscript{th} column, we also observe $m_i$ vertically resolved meteorological covariates $x_i^{(1)}, \ldots, x_i^{(m_i)}$ and their respective altitudes $h_i^{(1)} < \ldots < h_i^{(m_i)}$, such that $x_i^{(j)}\sim p(x|h_{i}^{(j)})$. We concatenate these observations into a dataset $\cD = \left\{\left(x_i^{(j)}, h_i^{(j)}\right)_{j=1}^{m_i}, \tau_i\right\}_{i=1}^n$ and denote $M = \sum_{i=1}^n m_i$ the total number of vertically resolved samples. The model description for the $i$\textsuperscript{th} column is summarized in Figure~\ref{fig:model-description}.

\begin{figure*}[t]
\begin{CatchyBox}{Model formulation for the $i$\textsuperscript{th} atmospheric column}
	    \begin{minipage}[h]{0.4\linewidth}
	    \strut
	    \newline
	    \textbf{Observation Model:}
        \begin{align*}
            \tau_i|\eta_i & \sim \cL\!\cN\left(\log\eta_i - \frac{\sigma^2}{2}, \sigma\right) \\
            \eta_i & = \int_0^H\varphi(x_i|h)\d h
        \end{align*}
	    \textbf{Prior:}
        \begin{align*}
            \varphi(x_i|h) & = \psi(f(x_i|h))e^{-h/L} \\
            f & \sim \operatorname{GP}(m, k)
        \end{align*}
	\end{minipage}
    \hfill
    \begin{minipage}[h]{.6\linewidth}
        \centering
        \begin{tabular}{cc}
        \toprule
             $\tau_i$ & Observed AOD \\
             $\cL\!\cN$ & Log-normal distribution \\
             $\eta_i, \sigma$ & Log-normal mean and scale parameters \\
             $\varphi$ & Prior for $b_\text{ext}$ \\
             $x_i|h$ & Input covariates at altitude $h$  \\
             $H$ & Atmospheric column height \\
             $\psi$ & Positive link function \\
             $L$ & Idealized profile heightscale parameter \\
             $f$ & GP prior with mean $m$ and kernel $k$ \\
        \bottomrule
        \end{tabular}
    \end{minipage}
\end{CatchyBox}
\caption{Observation model and prior formulation for the $i$\textsuperscript{th} atmospheric column.}
\label{fig:model-description}
\end{figure*}

Our objective is to use $\cD$ to learn the mapping $\varphi(x|h)$. Within the Bayesian framework, this objective is two-fold:
\begin{enumerate}
    \item[(A)] Update the prior placed on $\varphi$ with the observations from $\cD$. This corresponds to computing the posterior distribution of $\varphi$ given $\btau$.
    \item[(B)] Tune the model hyperparameters by maximising marginal log-likelihood $\log p(\btau)$. The hyperparameters include the log-normal scale parameter $\sigma$ and any parameter from the GP mean $m$ and kernel $k$.
\end{enumerate}

Regarding point (A), since $\varphi$ results from a transformation of GP $f$, we will rather focus on the posterior distribution of the GP directly for convenience. If $\f$ denotes a realization of $f$ at any input $x|h$, the posterior distribution of $f(x|h)$ given observations is denoted $p(\f|\btau)$. Having access to the posterior $p(\f|\btau)$ allows to compute the predictive means and variance of $\varphi(x|h)$ following
\begin{align}
    \EE[\varphi(x|h)|\btau] & = \int_\RR \psi(\f)e^{-h/L}p(\f|\btau)\d\f \label{eq:prior-posterior-mean}\\
    \operatorname{Var}(\varphi(x|h)|\btau) & = \EE[\varphi(x|h)^2|\btau] - \EE[\varphi(x|h)|\btau]^2\label{eq:prior-posterior-var}.
\end{align}
The above can be estimated with Monte Carlo by drawing samples from the posterior $p(\f|\btau)$. In Section~\ref{subsubsection:exponential-link} we will obtain a closed-form solution for the particular choice $\psi=\exp$.

Regarding point (B), the marginal log-likelihood $\log p(\btau)$ is unfortunately intractable. Indeed, let $\bfx = \begin{bmatrix}x_1^{(1)} & \hdots & x_n^{(m_n)}\end{bmatrix}^\top\in\cX^M$ denote the concatenation of all input entries from the dataset and let $\bbf\in\RR^M$ denote a realization of $f(\bfx)$. The marginal likelihood $p(\btau)$ can be expressed in terms of the observation model and prior distributions following
\begin{equation}
    p(\btau) = \int_{\RR^M} p(\btau|\bbf)p(\bbf)\d\bbf\enspace \enspace\text{ with }\enspace p(\btau|\bbf) = \prod_{i=1}^n p(\tau_i|\bbf_i),
\end{equation}
where $\bbf_i = \begin{bmatrix}\bbf_i^{(1)} & \hdots & \bbf_i^{(m_i)}\end{bmatrix}^\top\in\RR^{m_i}$ corresponds to the GP realization over the $i$\textsuperscript{th} column only. Because the observation model is log-normal, this integral is however intractable. To circumvent this intractability, we propose to tackle objective (B) by maximising a proxy of the marginal log-likelihood presented in Section~\ref{subsubsection:tuning}.

\subsubsection{A sparse variational approximation of the posterior}
We start by addressing objective (A). The predictive posterior distribution of interest is given by
\begin{equation}
    p(\f|\btau) = \frac{p(\btau|\f)p(\f)}{\int_\RR p(\btau|\f)p(\f)\d\f} .
\end{equation}
The integral denominator is not available in closed-form, making the posterior intractable. We propose to use a variational approximation scheme~\cite{tsitsias2009variational, matthews2016sparse, leibfried2020tutorial} to substitute this intractable inference problem with a tractable optimization problem. In addition, we make the variational approximation sparse~\cite{tsitsias2009variational} such that the model can scale to large amounts of data.

Let $\bfw = \begin{bmatrix} w_1 & \hdots & w_p\end{bmatrix}^\top\in\cX^p$ be a set of $p\ll M$ inducing locations over the space of inputs. Their evaluation by the GP follows a multivariate normal distribution $f(\bfw)\sim\cN(0, \bfK_{\bfw\bfw})$, where $\bfK_{\bfw\bfw} = k(\bfw, \bfw)$. We denote $\bfu=f(\bfw)\in\RR^p$ and refer to this vector as \emph{inducing variables}.

A $p$-dimensional parametric distribution is then set over these inducing variables. We choose this distribution as a multivariate normal defined by $q(\bfu) := \cN(\bfu|\bmu_\bfw, \bSigma_\bfw)$. $\bmu_\bfw, \bSigma_\bfw$ are called the \emph{variational parameters} and need to be tuned such that $q(\bfu)$ best approximates the true posterior $p(\bfu|\btau)$.

Once this is achieved, we take as an approximation to $p(\f|\btau)$ the variational posterior defined by $q(\f) := \int_{\RR^p} p(\f|\bfu)p(\bfu)\d\bfu$, which is given in closed-form by
\begin{align}
    q(\f) & = \cN(\f|\boldsymbol{\bar\mu}_{x|h}, \boldsymbol{\bar\Sigma}_{x|h}) \label{eq:qf-x|h}\\
    \boldsymbol{\bar\mu}_{x|h} & = k(x|h, \bw)\bfK_{\bfw\bfw}^{-1}\bmu_\bfw \\
    \boldsymbol{\bar\Sigma}_{x|h} & = k(x|h, x|h) - k(x|h, \bfw)\left(\bfK_{\bfw\bfw}^{-1} - \bfK_{\bfw\bfw}^{-1}\bSigma_\bfw\bfK_{\bfw\bfw}^{-1}\right)k(\bfw, x|h).
\end{align}

Naturally, (\ref{eq:qf-x|h}) can be extended to describe a variational posterior over multiple GP entries. Namely, let $\bfx^*\in\cX^D$ denote a vector of input entries. If $\bbf^*\in\RR^D$ denotes a realization of $f(\bfx^*)$, then the associated variational posterior is given by
\begin{align}
    q(\bbf^*) & = \cN(\bbf^*|\boldsymbol{\bar\mu}_{\bfx^*}, \boldsymbol{\bar\Sigma}_{\bfx^*}) \label{eq:qf}\\
    \boldsymbol{\bar\mu}_{\bfx^*} & = k(\bfx^*, \bw)\bfK_{\bfw\bfw}^{-1}\bmu_\bfw \label{eq:qf-mean}\\
    \boldsymbol{\bar\Sigma}_{\bfx^*} & = k(\bfx^*, \bfx^*) - k(\bfx^*, \bfw)\left(\bfK_{\bfw\bfw}^{-1} - \bfK_{\bfw\bfw}^{-1}\bSigma_\bfw\bfK_{\bfw\bfw}^{-1}\right)k(\bfw, \bfx^*). \label{eq:qf-covar}
\end{align}

The sparse nature of this approach becomes apparent in (\ref{eq:qf-mean}) and (\ref{eq:qf-covar}). Indeed, regardless of the number of samples we wish to evaluate the variational posterior over, we only need to invert a $p\times p$ matrix, incurring a $\cO(p^3)$ computational cost.

\subsubsection{Learning the variational parameters}\label{subsubsection:tuning}

As mentioned above, the variational parameters $\bmu_\bfw, \bSigma_\bfw$ need to be tuned such that $q(\bfu)$ best approximates the posterior $p(\bfu|\btau)$, which is intractable. This problem is casted as the maximisation of an objective called the evidence lower-bound (ELBO), given by
\begin{equation}\label{eq:elbo}
    \operatorname{ELBO}(q) = \EE_{q(\bbf)}[\log p(\btau|\bbf)] - \operatorname{KL}(q(\bfu)||p(\bfu)).
\end{equation}
The ELBO is a lower-bound to the marginal log-likelihood $\log p(\btau) \geq \operatorname{ELBO}(q)$. It can thus be used as a proxy of $\log p(\btau)$ to also tune the model hyperparameters and fulfill objective (B).

The second term in (\ref{eq:elbo}) is the Kullback-Leibler divergence between two multivariate normal distributions. It admits a closed-form expression and can be computed. The first term, on the other hand, is an expected log-likelihood under the variational posterior which cannot be analytically computed. It can be decomposed into column-wise terms
\begin{equation}\label{eq:ell}
    \EE_{q(\bbf)}[\log p(\btau|\bbf)]  = \sum_{i=1}^n \EE_{q(\bbf_i)}\left[\log\cL\!\cN\left(\tau_i|\log\eta_i - \frac{\sigma^2}{2}, \sigma\right)\right].
\end{equation}

To estimate (\ref{eq:ell}), we must first evaluate the mean parameter $\eta_i = \int_0^H\psi(f(x_i|h))e^{-h/L}\d h$ with the finite number of GP evaluations $\bbf_i$ we have access to. We propose to use the trapezoidal integration scheme given by
\begin{equation}\label{eq:trapez}
    \hat\eta_i = \sum_{j=1}^{m_i-1}\frac{\psi(\bbf_i^{(j+1)})e^{-h_i^{(j+1)}/L} - \psi(\bbf_i^{(j)})e^{-h_i^{(j)}/L}}{2}\left(h_i^{(j+1)} - h_i^{(j)}\right).
\end{equation}
While we choose the trapezoidal rule for simplicity, we note that alternative finite-sample integration schemes can be chosen here in accordance with the needs.

Second, because of the log-normal observation model, the expected log-likelihood remains intractable. To estimate it, while allowing backpropagation through the variational parameters, we use a reparametrization trick~\cite{kingma2015variational}. Namely, we sample $\beps_i\sim \cN(0, \bI_{m_i})$ and compute $\bbf_i = \boldsymbol{\bar\mu}_i + \boldsymbol{\bar\Sigma}_i^{1/2}\beps_i$, where $\boldsymbol{\bar\mu}_i, \boldsymbol{\bar\Sigma}_i$ are the variational posterior parameters for the $i$\textsuperscript{th} column and are obtained by application of (\ref{eq:qf-mean} and (\ref{eq:qf-covar}) over the predictors of the $i$\textsuperscript{th} column. The resulting GP sample $\bbf_i$ is then used to estimate the mean parameter $\hat\eta_i$ following (\ref{eq:trapez}) and we can approximate the expected log-likelihood with its one-sample estimate
\begin{equation}
    \EE_{q(\bbf_i)}\left[\log\cL\!\cN\left(\tau_i|\log\eta_i - \frac{\sigma^2}{2}, \sigma\right)\right] \approx \log\cL\!\cN\left(\tau_i|\log\hat\eta_i - \frac{\sigma^2}{2}, \sigma\right).
\end{equation}

This method allows to estimate the ELBO objective, which in turn can be maximised with respect to the variational parameters $\bmu_\bfw, \bSigma_\bfw$ using a stochastic gradient approach for example. As mentioned above, the model hyperparameters can also be tuned jointly with this objective, hence fulfilling objective (B). These include the log-normal scale $\sigma$ or the kernel $k$ hyperparameters (e.g.\ variances and lengthscales), with an option to parametrize these kernels using feature maps given by deep neural networks~\cite{law2019hyperparameter}. As it is standard in sparse variational GPs~\cite{tsitsias2009variational}, we will also learn the inducing locations $\bfw$.

\section{Experimental Setup}\label{section:experimental-setup}

Our motivating application is to reconstruct aerosol vertical profiles from remote-sensing AOD observations. However, to validate the proposed methodology, we also need to observe extinction coefficient profiles. While the latter can be collected from vertically resolved remote sensing instruments~\cite{winker2013global, holben1998aeronet}, the collocation of measurements from different devices raises non-trivial questions regarding the self-consistency of the collocated dataset --- and thus the validation procedure. In contrast, climate models are self-consistent and offer readily available simulations of gridded AOD and extinction coefficient profiles. We hence propose to evaluate our model using observations from a global climate model, the ECHAM-HAM global aerosol-climate model simulation~\cite{stier2005aerosol, stier2007aerosol, zhang2012global}.

In this section, we first describe the ECHAM-HAM dataset used in the experiments. We then outline the model setup and experimental procedure. Finally, we present the evaluation metrics used to quantify the quality of the reconstructed vertical profiles.

\subsection{Dataset and Experiment}

\subsubsection{The ECHAM-HAM dataset} The aerosol-climate model ECHAM-HAM is a self-consistent global climate model of aerosol radiative properties and CCN which demonstrates excellent agreement with AOD measurements from ground-based sun-photometers and satellite retrievals~\cite{stier2005aerosol}. It computes the evolution of log-normal aerosol mass and number modes --- for species sulfates, black carbon, organic carbon, sea salt and mineral dust --- by taking into account physical and chemical particle processes. The simulation used includes aerosols optical properties, aerosols tracers and meteorological variables at a 1.8$^\circ\times$1.8$^\circ$ horizontal resolution on a Gaussian grid and over 31 levels of vertical resolution and for 8 regularly spaced time steps over a day (06/06/2008). The resulting dataset counts $147\,456$ atmospheric columns used as training points. Table~\ref{table:dataset-dims} provides its detailed dimensions.
\begin{table}[t]
\centering
\begin{tabular}{rlcc}\toprule
                          & Name  & Notation & Dimensions
                          \\\midrule
\multirow{4}{*}{\textit{Predictors}} 
                          &  Temperature & $T$ & ($t$, lat, lon, lev)\\
                          &  Pressure & $P$ & ($t$, lat, lon, lev)\\
                          &  Relative humidity & RH & ($t$, lat, lon, lev)\\
                          &  Vertical velocity & $\omega$ & ($t$, lat, lon, lev)\\
                          \midrule
\multirow{1}{*}{\textit{Response}} 
                            & AOD 550nm & $\tau$ &  ($t$, lat, lon) \\
                          \midrule
\multirow{1}{*}{\textit{Groundtruth}} 
                            & Extinction coefficient 533nm & $b_\text{ext}$ &  ($t$, lat, lon, lev) \\
                          \bottomrule
\end{tabular}
\caption{Gridded variables from ECHAM-HAM simulation data. The grid includes 8 time steps ($t$), 96 latitude levels (lat), 192 longitude levels (lon) and 31 vertical pressure levels (lev) --- which is a proxy for $h$. Our objective is to vertically disaggregate the response $\tau$ using the vertically resolved predictors ($T, P, \text{RH}, \omega$) and spatiotemporal columns locations ($t$, lat, lon).}
\label{table:dataset-dims}
\end{table}

\subsubsection{Experiment objective} In this experiment, we apply our model to the vertical disaggregation of the AOD simulated by ECHAM-HAM. We use vertically resolved predictors and compare our prediction to the extinction coefficient simulated by ECHAM-HAM.

\subsubsection{Choice of variables}
 We use the AOD at 550nm as the response variable to vertically disaggregate and the extinction coefficient at 533nm as the groundtruth variable to evaluate our predictions against. To select the vertically resolved variables used as predictors, we limit ourselves to standard meteorological variables which could be obtained from reanalysis data: $T$, $P$, RH and $\omega$. Indeed, while aerosol satellite imagery do not provide any vertically resolved measurements of meteorological variables, the latter can be reliably extracted on different pressure-levels from reanalysis data or from atmospheric sounders. The input variable writes $x = (t, \text{lat}, \text{lon}, T, P, \text{RH}, \omega)$.

Predictors and AOD are standardized and meteorological predictors are transformed to follow approximately normal distributions using a rank-based inverse normal mapping.

\subsection{Model Setup and Tuning}

\subsubsection{Choosing the idealized profile heightscale $L$}

A simple choice for the idealized profile heightscale is the boundary layer upper altitude, typically taken at $2\,km$. This is the value assumed by MISR, VIIRS and MODIS C6\_DT AOD retrieval algorithms~\cite{kahn2015analysis, laszlo2016eps, levy2007second}. In our experiment, we also choose to set $L=2\,km$.

\subsubsection{Sparse variational GP setup}

We set the GP with a zero mean function $m=0$ and the following kernel
\begin{align}
    k(x|h, x'|h') & =  \gamma_1 C_{3/2}(t, t')C_{3/2}(\text{lat}, \text{lat}')C_{3/2}(\text{lon}, \text{lon}')  \label{eq:spatiotemporal-kernel}\\
    & + \gamma_2 C_{1/2}\left([T, P, \omega, \text{RH}]|h, [T', P', \omega', \text{RH}']|h'\right), \label{eq:meteo-kernel}
\end{align}
where $C_\nu$ is the Matérn-$\nu$ covariance function with automatic relevance determination --- i.e.\ independent lengthscale parameters for each entry. This is simply the sum of a spatiotemporal kernel (with variance $\gamma_1$ and lengthscales $\ell_t, \ell_\text{lat}, \ell_\text{lon}$) and a kernel on meteorological predictors (with variance $\gamma_2$ and lengthscales $\ell_T, \ell_P, \ell_\text{RH}, \ell_\omega$). The Matérn family has been widely used to work with spatial data~\cite{stein1999interpolation}. This choice of Matérn order $\nu$ guarantees that the GP is continuous with respect to meteorological predictors and continously differentiable with respect to time and space. Explicit expressions of the covariance functions are given in Appendix~\ref{appendix:matern-covariance}.

The spatiotemporal kernel (\ref{eq:spatiotemporal-kernel}) makes the predicted extinction smooth across time and space. Its product structure ensures that for distant times ($t, t'$), latitudes (lat, lat$'$) or longitudes (lon, lon$'$), the spatiotemporal covariance vanishes to zero. Hence, the covariance of predictions distant in space or time will disregard spatiotemporal predictors. This prevents overfitting over spatial and temporal predictors.

Independently from the spatiotemporal kernel, the kernel on meteorological predictors (\ref{eq:meteo-kernel}) introduces covariance between predicted extinctions if meteorological predictors are close, even for distant times and locations. This kernel captures how meteorological predictors modulate extinction. In practice, we find that fixing $\gamma_2=1$ helps to prevent vanishing kernel signal when tuning the hyperparameters.

We use $p=60$ inducing locations $\bfw$ as an arbitrary choice (fewer or more inducing locations can be used following needs). They are initialized by randomly drawing samples at boundary layer altitude from $\bfx$. The variational distribution $q(\bfu)= \cN(\bfu|\bmu_\bfw, \bSigma_\bfw)$ is initialized with $\bmu_\bfw = 0$ and $\bSigma_\bfw = \bI_p$.

\subsubsection{Mini-batch tuning}
We denote $\Theta = \left\{\bmu_\bfw, \bSigma_\bfw, \bfw, \sigma, \gamma_1, \gamma_2, \ell_t, \ell_\text{lat}, \ell_\text{lon}, \ell_T, \ell_P, \ell_\text{RH}, \ell_\omega\right\}$ the joint set of variational parameters and model hyperparameters. We maximise the ELBO objective from (\ref{eq:elbo}) with respect to $\Theta$ using Adam optimiser~\cite{kingma2015adam}.

Since the dataset includes $8\times 96\times 192 = 147\,456$ atmospheric columns, evaluating the ELBO for all columns at each optimisation step is computationally prohibitive. Instead, we use a stochastic gradient approach and sample mini-batches of 64 columns over which we compute the ELBO and take gradient steps.

\subsection{Inference Procedure}

\subsubsection{Exponential link}\label{subsubsection:exponential-link}

We choose an exponential link function ($\psi=\exp$) to warp the GP. This particular choice is convenient as it makes the prior $\varphi(x|h)$ a log-normal random variable given by
\begin{equation}\label{eq:prior-with-exp}
    \varphi(x|h) = e^{f(x|h) - h/L}\sim \cL\!\cN\left(m(x|h) - h/L, k(x|h, x|h)^{\nicefrac{1}{2}}\right).
\end{equation}
A direct consequence is that we can obtain analytical expressions of its moments, such as the posterior mean from (\ref{eq:prior-posterior-mean}) now given by
\begin{equation}
    \EE[\varphi(x|h)|\btau] = \exp\left(m(x|h) - \frac{h}{L} + \frac{1}{2}k(x|h, x|h)\right).
\end{equation}
We can also express its quantiles in closed-form without resorting to estimation procedures. This is useful to compute confidence regions.

\subsubsection{Rescaling the predicted profiles}\label{subsubsection:rescaling-profiles}
The posterior distribution $p(\bbf|\btau)$ updates the GP with the information that the mapping $\varphi$ should (in expectation) vertically integrate to the observed AOD $\btau$. While instructive, this is a weak constraint on the prior and does not guarantee that the model will effectively integrate to the AOD of the atmospheric column.

\begin{figure}[t]
    \centering
    \includegraphics[width=\textwidth]{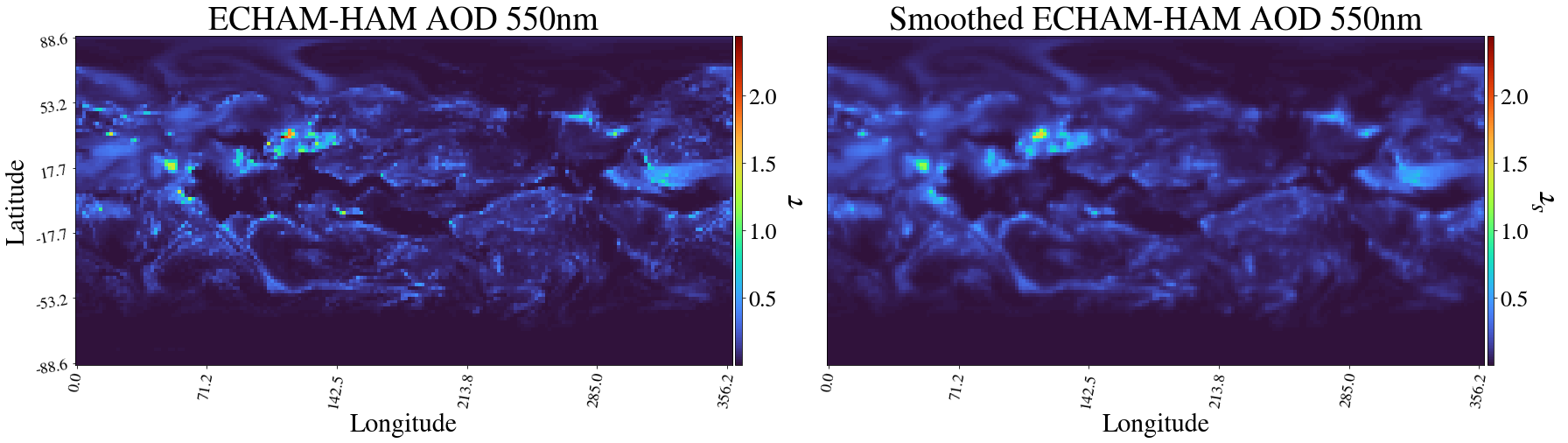}
    \caption{\textbf{Left:} ECHAM-HAM 550nm AOD. \textbf{Right:} Spatially smoothed ECHAM-HAM 550nm AOD}
    \label{fig:aod-vs-smoothed-aod}
\end{figure}

To get a stronger enforcement of column-integrated values, we could simply rescale the model against observed AOD. Observations $\tau_1, \ldots, \tau_n$ are however corrupted with high-frequency noise and might be quite different from the actual AOD. To filter out this noise, we introduce a spatially smoothed version of the observations denoted $^s\!\tau_1, \ldots, ^s\!\tau_n$. We obtain it by applying a Gaussian smoothing filter across latitude and longitude to the observations $\tau_1, \ldots, \tau_n$. The resulting fields are displayed in Figure~\ref{fig:aod-vs-smoothed-aod}.

At inference time, we substitute the posterior with its rescaled version given by
\begin{equation}
    ^s\!\varphi(x_i|h)|\btau = \frac{^s\!\tau_i}{\int_0^H \EE[\varphi(x_i|h)|\btau]\d h} \varphi(x_i|h)|\btau
\end{equation}
for the $i$\textsuperscript{th} column. When choosing $\psi=\exp$, this simply corresponds to shifting the location of (\ref{eq:prior-with-exp}) by $\log\,\!^s\!\tau_i  - \log \int_0^H \EE[\varphi(x_i|h)|\btau]\d h$. We approximate the integral using a trapezoidal scheme.

\subsubsection{From $^s\!\varphi(x|h)$ to $b_\text{ext}(h)$}\label{subsubsection:sigmaext-calibration}

The final stage of inference is to estimate the posterior distribution of $b_\text{ext}$ following the observation model (\ref{eq:bext-observation-model}). Because the chosen observation model conserves the mean, the posterior mean of $b_\text{ext}(h)$ is the same as the posterior mean of $^s\!\varphi(x|h)$. Higher-order moments however cannot be obtained analytically and will require estimation by sampling from $b_\text{ext}(h)|\btau$. This can be achieved by sampling from $^s\!\varphi(x|h)|\btau$ and then plugging the samples in the observation model (\ref{eq:bext-observation-model}) to sample from $b_\text{ext}(h)|^s\!\varphi(x|h)$.

The observation model scale parameter $\sigma_\text{ext}$ needs to be calibrated against observational noise. We set $\sigma_\text{ext}$ to the value that minimises the integrated calibration index (ICI) given by
\begin{equation}\label{eq:ICI}
    \operatorname{ICI} = \int_0^1 |N(\alpha) - (1-\alpha)|\d\alpha,
\end{equation}
where $N(\alpha)$ is the percentage of ECHAM-HAM extinction coefficient observations that fall within the $(1-\alpha)$ credible interval of the distribution of $b_\text{ext}|\btau$. The ICI is evaluated against $n_\text{calib} = 200$ random atmospheric columns from the dataset. These columns form a separate calibration set which is solely used to calibrate $\sigma_\text{ext}$. This separate calibration set is negligible in size in comparison with the $147\,456$ atmospheric columns available in the simulation.

\subsection{Evaluation procedure}

We compare the predicted extinction coefficient profile $b_\text{ext}|\btau$ to the extinction coefficient from ECHAM-HAM simulations.

\subsubsection{Baseline}

We use as a comparative baseline the idealized exponential profile. For a fair comparison, we rescale the baseline profiles such that their integrated values match the AOD, following the procedure of Section~\ref{subsubsection:rescaling-profiles}. We also postulate a log-normal observation model for the extinction coefficient and calibrate its scale parameter against a separate calibration set as described in Section~\ref{subsubsection:sigmaext-calibration}. This allows to perform a probabilistic evaluation of the extinction coefficient profiles predicted with the idealized baseline.

\subsubsection{Evaluation}

We use two kind of metrics to quantify the quality of the predicted vertical profiles: deterministic metrics, which compare only the posterior mean prediction to the groundtruth ECHAM-HAM extinction coefficient profiles, and probabilistic metrics which evaluate the entire posterior probability distribution against the extinction coefficients from ECHAM-HAM simulation and thus also assess the quality of the uncertainty quantification in the posteriors. The metrics used are outlined in Table~\ref{table:metrics}.

\begin{table}[t]
\centering
\caption{Evaluation metrics; Deterministic metrics compare the predicted posterior mean $\EE[b_\text{ext}|\btau]$ to the ECHAM-HAM extinction coefficient; Probabilistic metrics evaluate the complete predicted posterior probability distribution of $b_\text{ext}|\btau$ against ECHAM-HAM extinction coefficient.}
\begin{tabular}{rlll}
\toprule
                               & Metric  & Description    &   Best when                                                                                     \\
\midrule
\multirow{5}{*}{\textit{Deterministic}} & RMSE    & Root mean square error   & close to 0                                                                                \\
                               & MAE  & Mean absolute error     &   close to 0                                                                      \\
                               & Corr    & Pearson correlation &  close to 100\%                                                                \\
                               & Bias    & Mean bias     &   close to 0                                                                           \\
                               & Bias98  & Bias in the 98th percentile &   close to 0                                                                  \\
\midrule
\multirow{3}{*}{\textit{Probabilistic}} & ELBO    & Evidence lower-bound of groundtruth $b_\text{ext}$ & higher \\
                               & Calib95 & 95\% calibration score, i.e.\ $N(\alpha=0.05)$ & close to 95\%                                                                 \\
                               & ICI     & Integrated calibration index --- see (\ref{eq:ICI}) & close to 0 \\
\bottomrule
\end{tabular}
\label{table:metrics}
\end{table}

For each metric, we compute scores for pixels along the entire atmospheric columns and for pixels lying within the boundary layer only ($<2\,km$). Scores are averaged across all pixels.

\section{Results}\label{section:results}

\subsection{Reconstructed vertical profiles}

As reported in Table~\ref{table:scores}, the posterior mean profile arising from the proposed method outperforms the idealized exponential baseline for all deterministic metrics. Evaluating over the entire column consistently yields better scores than over the boundary layer only. This is to be expected as the extinction coefficient outside the boundary layer tends to vanish to zero and most of the variability happens within the boundary layer.

\begin{table}[t!]
\centering
\caption{Comparison of our method to an idealized exponential baseline for deterministic and probabilistic metrics; \enquote{Entire column} means scores are computed and averaged for every altitude levels; \enquote{Boundary layer} means scores are computed and averaged for altitude levels of the boundary level only ($<2\,km$); runs with our method are averaged over 5 seeds; we report 1 standard deviation.}
\begin{tabular}{lrccccc}
\toprule
    \textit{Region}     &   Method   & RMSE \scriptsize{(10\textsuperscript{-5})} & MAE \scriptsize{(10\textsuperscript{-6})} &  Corr \scriptsize{(\%)} &  Bias \scriptsize{(10\textsuperscript{-6})} &  Bias98 \scriptsize{(10\textsuperscript{-5})} \\
\midrule
\multirow{2}{*}{\thead{\textit{Entire} \\ \textit{column}}}      & Our method   & \textbf{3.29}\footnotesize{$\pm$0.02} & \textbf{5.31}\footnotesize{$\pm$0.07} &  \textbf{70.9}\footnotesize{$\pm$0.4} & \textbf{-0.167}\footnotesize{$\pm$0.105} & \textbf{-0.646}\footnotesize{$\pm$0.151}\\
                          & Idealized & 4.10  & 6.65 & 51.2 & -2.40  & -4.08  \\

\midrule
\multirow{2}{*}{\thead{\textit{Boundary} \\ \textit{layer}}} & Our method  & \textbf{6.06}\footnotesize{$\pm$0.03} & \textbf{15.2}\footnotesize{$\pm$0.3} &  \textbf{69.8}\footnotesize{$\pm$0.5} & \textbf{-1.25}\footnotesize{$\pm$0.45} & \textbf{-4.64}\footnotesize{$\pm$0.32}  \\
                          & Idealized & 7.55  & 16.8 & 53.6 & -12.9  & -11.7  \\
\bottomrule
\end{tabular}

\strut

\begin{tabular}{lrccc}
\toprule
    \textit{Region}     &   Method   & ELBO &  Calib95 \scriptsize{(\%)} &  ICI \scriptsize{(10\textsuperscript{-2})} \\
\midrule
\multirow{2}{*}{\thead{\textit{Entire} \\ \textit{column}}}      & Our method   & \textbf{13.1}\footnotesize{$\pm$0.1} &  \textbf{94.9}\footnotesize{$\pm$0.1} & 5.29\footnotesize{$\pm$0.59} \\
                          & Idealized & \textbf{13.1} & 96.0 & \textbf{5.05}  \\

\midrule
\multirow{2}{*}{\thead{\textit{Boundary} \\ \textit{layer}}} & Our method  & \textbf{10.6}\footnotesize{$\pm$0.1} &  98.8\footnotesize{$\pm$0.1} & \textbf{8.27}\footnotesize{$\pm$0.29}  \\
                          & Idealized & 10.2 & \textbf{93.5} & 19.1 \\
\bottomrule
\end{tabular}
\label{table:scores}
\end{table}

The RMSE and MAE are improved with our method. The RMSE and MAE gap with the baseline slightly widens if we look at the boundary layer only. This suggests our method is better than the idealized exponential profile at predicting the extinction coefficient specifically within the boundary layer. Our method improves the mean bias by an order of magnitude both for the entire column and the boundary layer. Similarly, the bias in the 98th percentile is consistently improved by an order of magnitude, suggesting our method better captures tail behaviour of the extinction profile.

The ELBO\footnote{the ELBO computed for the idealized baseline corresponds exactly to the marginal log-likelihood whereas the ELBO computed for our method is a lower-bound to the marginal log-likelihood, which could possibly be greater.} and ICI are comparable for both methods when computed over the entire column. For the boundary layer however, our method outperforms the baseline in ELBO and ICI, with a significant difference for the ICI. This suggests that, in the boundary layer, the predicted posterior probability distribution with our method is a sounder fit to the ECHAM-HAM extinction coefficient profiles. The 95\% calibration scores are close to 95\% and comparable for both methods --- with a slight edge for the baseline in the boundary layer.

\begin{figure}[H]
    \centering
    \includegraphics[width=0.91\textwidth]{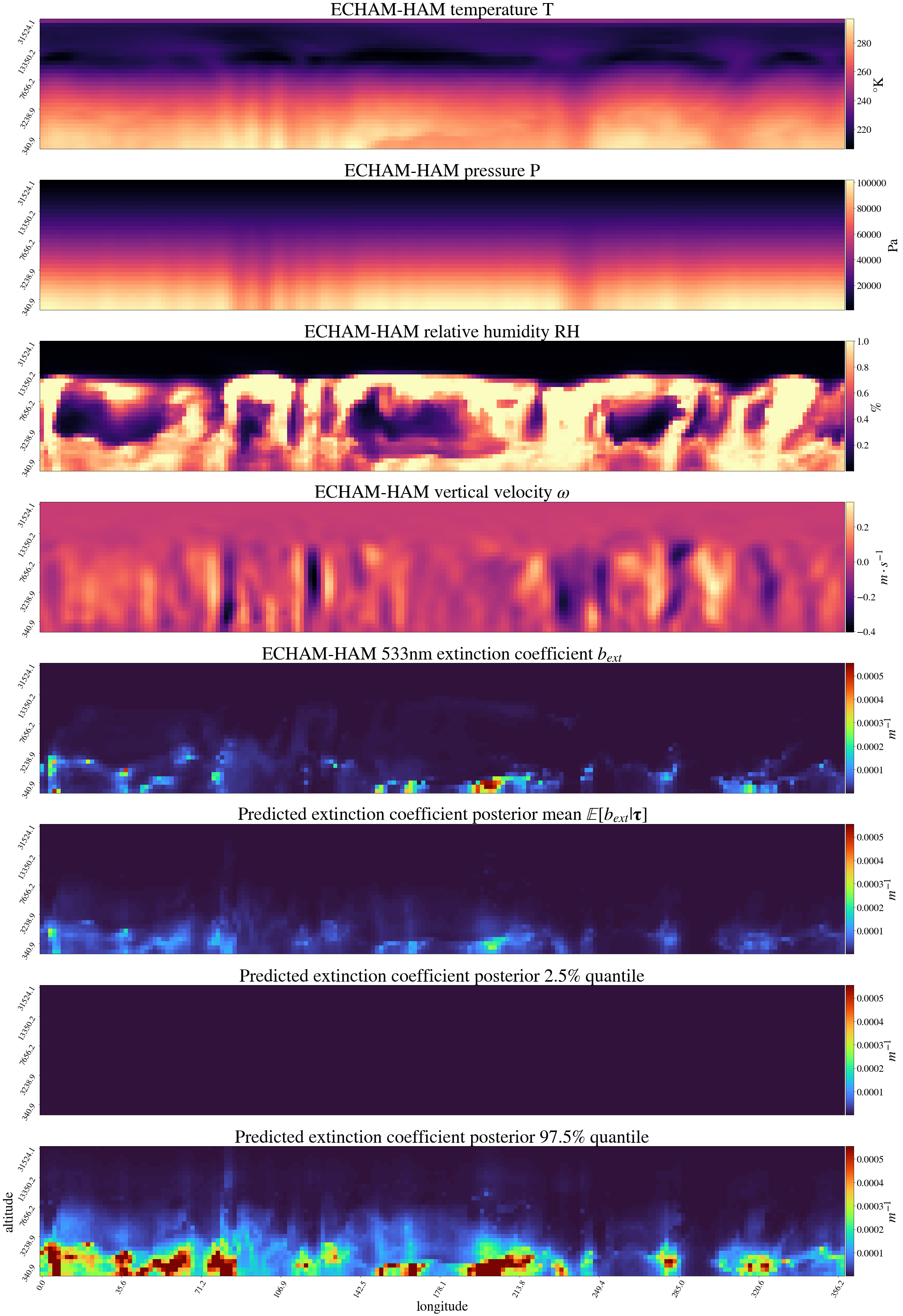}
    \caption{Vertical slices at latitude 51.29$^\circ$ of meteorological predictors ($T, P, \text{RH}, \omega$), groundtruth extinction coefficient, predicted extinction coefficient posterior mean, 2.5\% and 97.5\% quantiles of the predicted extinction coefficient posterior distribution}
    \label{fig:slices}
\end{figure}

\begin{figure}[t]
    \centering
    \includegraphics[width=0.91\textwidth]{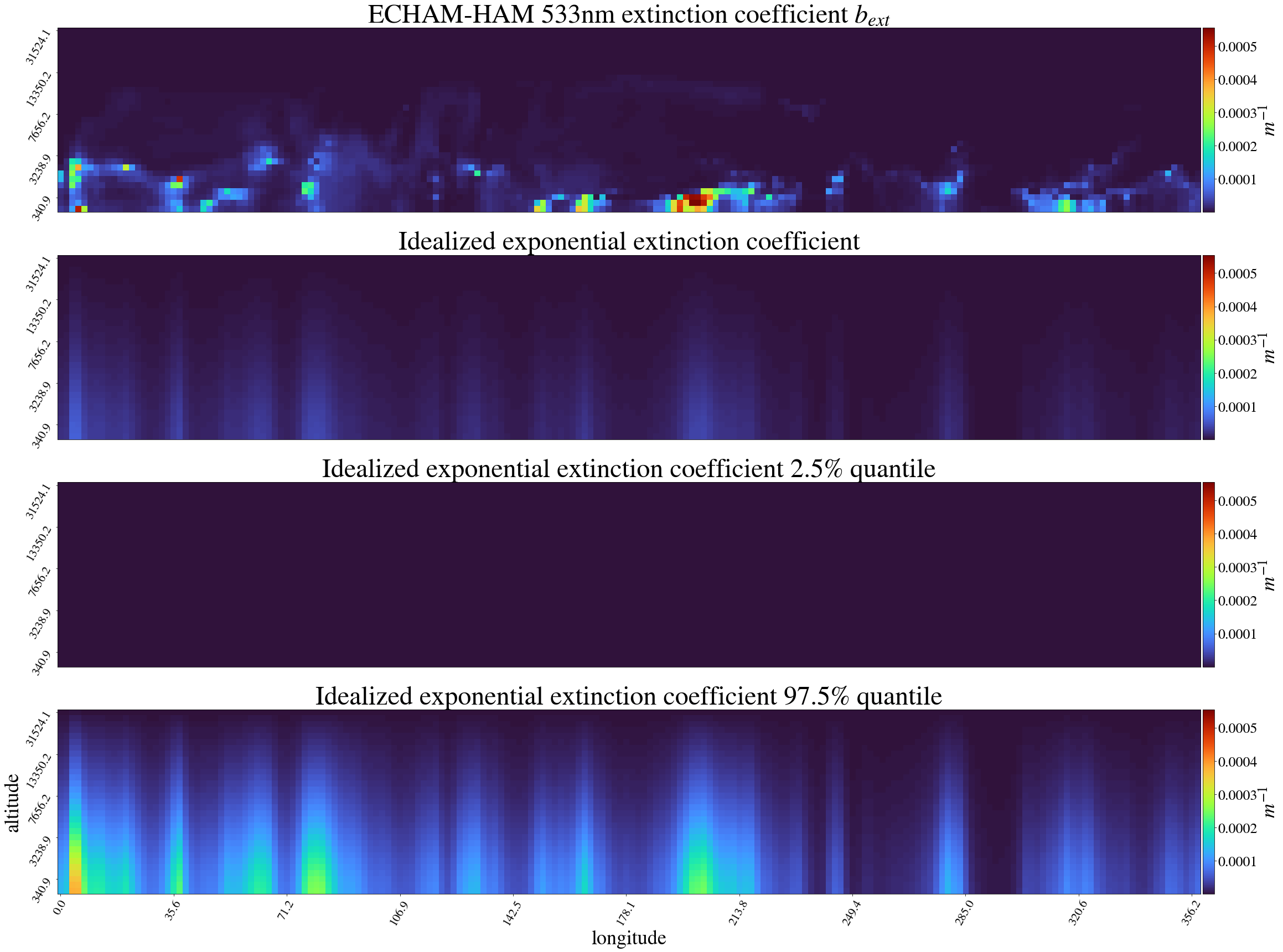}
    \caption{Vertical slices at latitude 51.29$^\circ$ of groundtruth extinction coefficient, idealized exponential extinction coefficient, 2.5\% and 97.5\% quantiles of the idealized exponential extinction coefficient distribution}
    \label{fig:ideal-slices}
\end{figure}

Figure~\ref{fig:slices} displays slices at fixed latitude of the vertically resolved predictors used, the ECHAM-HAM extinction coefficient and the predicted extinction coefficient. For comparison, we also display in Figure~\ref{fig:ideal-slices} the prediction at the same latitude with the idealized exponential baseline. Additional sliced predictions are provided in Appendix~\ref{appendix:additional-plots} for different latitudes.

We observe that our predicted mean profile is able to reconstruct extinction patterns that are visually very similar the the groundtruth extinction coefficient. In comparison with the idealized exponential baseline, the extinction profiles predicted with our method look much more realistic. This is encouraging given the only aerosol optical property used is the AOD. In particular, since aerosols water uptake is related to relative humidity, we observe a good capacity to recover patterns corresponding to extinction due to aerosol swelling in the boundary layer. We also observe that tail extinction coefficient values are consistently captured within the 95\% confidence region of the posterior distribution.

The predicted mean profile however fails to reproduce some extinction patterns, such as the ones depicted for longitudes $<100^\circ$ in Figure~\ref{fig:slices}. We conjecture these are extinction patterns imputable to aerosols mass concentration, particles size and radiative properties. These are harder properties which, unlike extinction due to swelling, cannot be fully characterized by relative humidity, temperature, pressure and updraft.

The predictions also tend to be smoother than the groundtruth extinction coefficient. This is a regularizing property of the prior which leads to overestimation of the extinction coefficient, in particular in regions with very low extinction between and above extinction pockets. For example, around longitude $200^\circ$, the ECHAM-HAM extinction in Figure~\ref{fig:slices} displays an extinction pocket with a sharp limit around altitude $680\,m$, above which extinction is virtually absent. In contrast, our predicted posterior mean is more diffuse and tends to lightly spread out above $680\,m$.

\begin{figure}[t]
    \centering
    \includegraphics[width=0.8\textwidth]{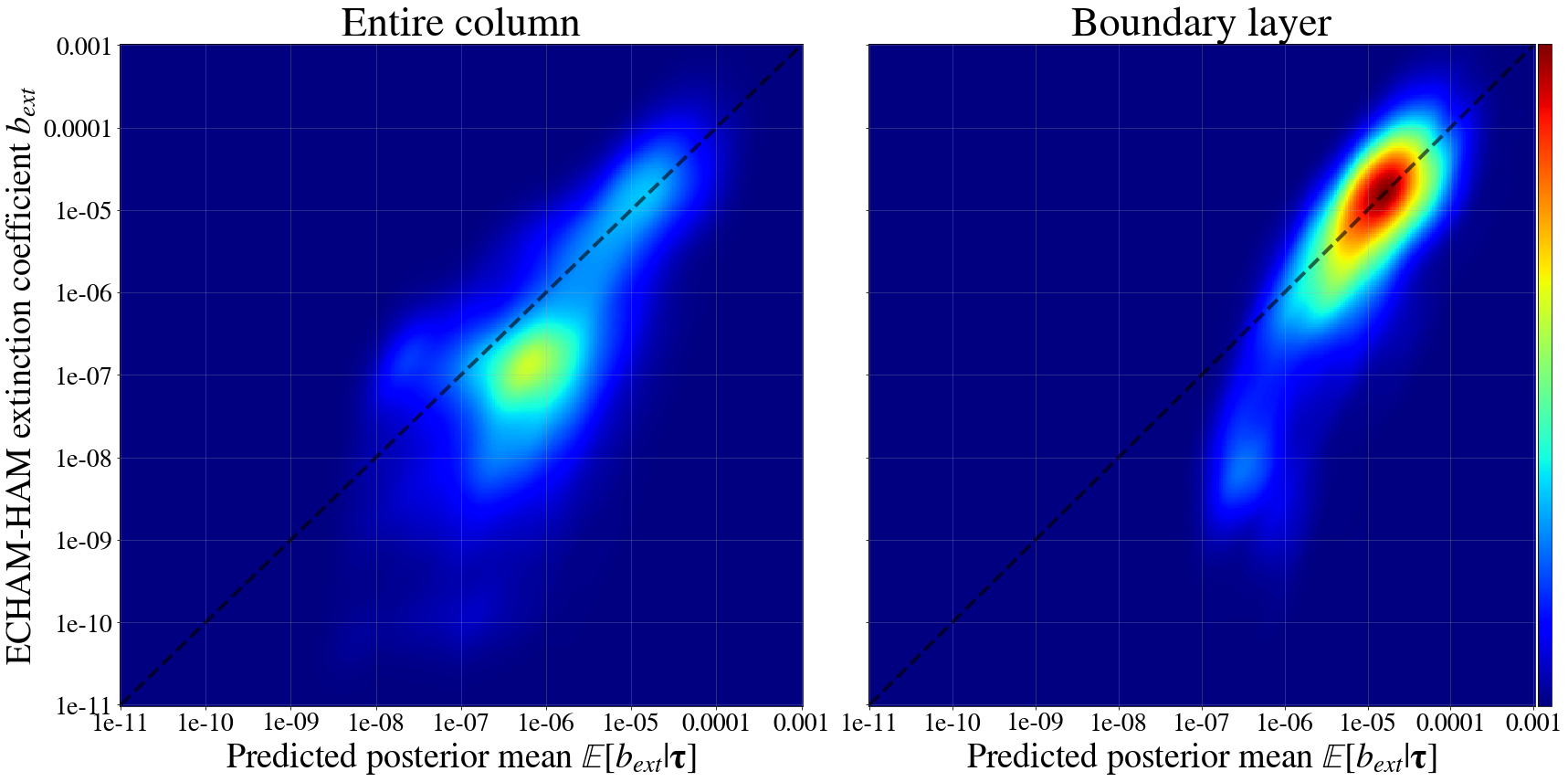}
    \caption{Density plots of groundtruth extinction coefficient values against predicted posterior mean extinction coefficient; \textbf{Left:} plotted for the entire column; \textbf{Right:} plotted for the boundary layer only; density plots are computed on a random subset on a random subset of 1000 samples drawn for the entire column (left) and in the boudary layer (right)}
    \label{fig:density-plots}
\end{figure}

The density plots in Figure~\ref{fig:density-plots} support this observation as a large mass of extinction coefficient around $10^{-7}\,m^{-1}$ tends to be overestimated by an order of magnitude. This trend is however significantly reduced if we focus on the boundary layer only where most of the mass lies around the axis $y=x$. This suggests that most of the overestimation happens above the boundary layer for low extinction ($<10^{-6}\,m^{-1}$). Within the boundary layer however, the mean predictions are reasonably aligned with the groundtruth for high extinction coefficient values ($>10^{-5}\,m^{-1}$). Low extinction coefficient values are significantly less dominant in the boundary layer, but also tend to be overestimated. We attribute this behavior to the over-smoothing of the GP prior, which is prone to overestimation between extinction pockets.

\subsection{Feature importance}

\begin{figure}[t]
    \centering
    \includegraphics[width=\textwidth]{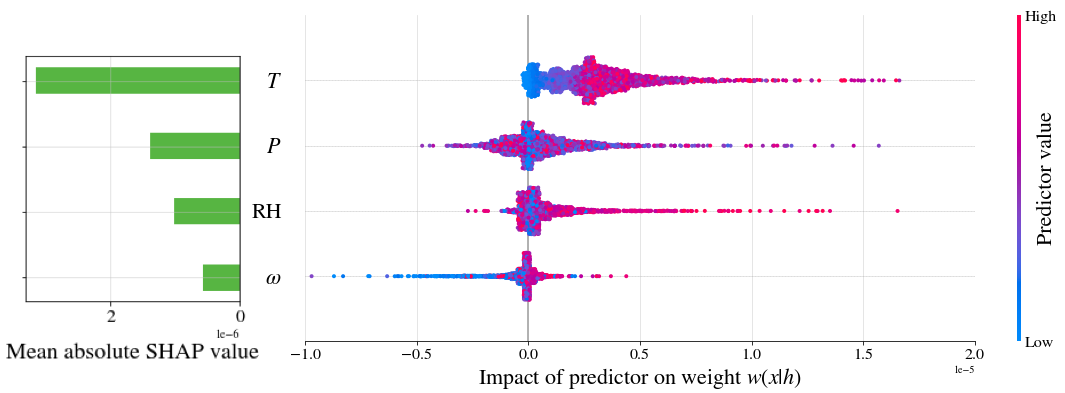}
    \caption{\textbf{Left:} Mean absolute Shapley values of meteorological predictors in the boundary layer; \textbf{Right:} Beeswarm plot of Shapley values of individual predictions in the boundary layer for each meteorological predictor; red (blue) dots indicate a sample where the predictor value lies on the right (left) tail of its distribution; Shapley values are computed on a random subset of 2000 samples in the boundary layer}
    \label{fig:shap}
\end{figure}

To understand the contribution of each meteorological predictor ($T$, $P$, RH, $\omega$) to extinction coefficient predictions, we study feature importance using Shapley values (SVs)~\cite{shapley1953value}. Originally introduced as a game-theoretic concept, SVs have been widely adopted in machine learning to design local feature importance explanation models~\cite{vstrumbelj2014explaining, lundberg2018consistent, ghorbani2019data, lundberg2017unified}. With SVs, we can estimate for each individual prediction how much each predictor contributed, and hence explain predictions. In our study, we use the KernelSHAP SV-based explanation model~\cite{lundberg2017unified}.

Since the meteorological predictors only modulate predictions through the weighting function, we only study the impact meteorological predictors have on the weight $w(x|h)$ predicted for the posterior mean prediction. We also focus on samples within the boundary layer since this is where most of the variability happens. Figure~\ref{fig:shap} displays the mean absolute SV obtained for each meteorological predictor and the contribution each predictor has on a randomly selected set of individual predictions.

We observe that in absolute mean, temperature is the factor that most influences predictions of the weighting function in the boundary layer. This is sensible as temperature is a smooth spatial field that gradually decreases with altitude, making $T$ characteristic of the altitude level. Since the altitude typically correlates with extinction values, $T$ becomes an informative proxy of extinction. This is corroborated by the plot of contributions to individual predictions: greater temperature (red) --- lower altitudes --- yield greater impact on the weight and hence greater extinction prediction while lower temperature (blue) --- higher altitudes --- yield lower impact on the weight and hence lower extinction prediction. It is interesting to notice that while pressure is also a proxy of altitude just like temperature, its mean absolute contribution is lower and the individual contribution plot does not display a similar trend. This suggests the information conveyed by pressure might be redundant with the information conveyed by temperature.

While the mean absolute contribution of relative humidity seems marginal compared to temperature, we notice in individual contributions that when RH positively impacts predictions, it consistently corresponds to locations with high RH. This supports the observation that our model captures extinction due to aerosol water uptake. Similarly, negative contributions of $\omega$ to the weight consistently correspond to low $\omega$. This is consistent with intuition as low updrafts lead to low humidification of aerosols, and a reduction in water content in aerosol decreases extinction due to aerosol water uptake.

\section{Discussion and Future Directions}\label{section:discussion}

In this work, we introduce a GP-based methodology to vertically disaggregate the AOD using simple vertically resolved meteorological predictors such as temperature, pressure or relative humidity. Our approach emphasises uncertainty quantification using a Bayesian formalism. A successful application of our methodology to the vertical disaggregation of ECHAM-HAM simulated AOD is demonstrated. Our model outperforms an idealized baseline and displays capacity to recover realistic extinction patterns, in particular for extinction patterns arising from aerosol swelling in the boundary layer.

The simplicity of the explicit modelling assumptions grants better control and interpretability over the model and makes the choice of analysis strategy less subjective. While such simplicity can never account for the complex phenomena underpinning ACIs, this is balanced by a particular emphasis on quantification of epistemic uncertainty through a principled Bayesian formalism.

Naturally, the modelling assumptions can be adapted to reflect different modelling choices. For example, where more relevant vertically resolved quantities can be obtained, they can seamlessly be incorporated into the predictors. The idealized exponential component of the prior can be replaced by an idealized Gaussian profile to reflect the assumptions made by different AOD retrieval algorithms~\cite{li2020impact, wu2017impact}. The kernel design is also flexible and allows the user to specify a covariance structure that best accounts for pairwise dependencies of the predictors. Similarly, the positive link function, integration scheme or number of inducing locations can be modified to fit specific needs.

Experiments demonstrate the proposed model is able to realistically reproduce vertical structures of extinction. In particular --- for the chosen set of meteorological predictors --- we are able to reliably predict extinction due to water vapour in the boundary layer. Because of the smoothness bestowed by the GP prior, the model tends to overestimate low extinction above the boundary layer and between extinction packets in the boundary layer. It is conjectured that the remaining unexplained extinctions patterns can be attributed to aerosols mass concentration, particles size and radiative properties, which are more challenging to model and would require additional vertically resolved covariates. This constitutes an important area for future work.

Regarding methodology, several extensions remain to be explored. For example, while we only consider aerosol extinction at a single wavelength, the AOD is in general retrieved for multiple wavelengths ranging from $470\,nm$ to $870\,nm$~\cite{remer2005modis}. By making the GP a function of the wavelength $f(\lambda, x|h)$, it should be possible to leverage multiple AOD observations by introducing a notion of functional smoothness across the electromagnetic spectrum. Another exciting direction would be to allow the use of unmatched AOD and vertically resolved predictors. This can be achieved by adapting the work of \citet{chau2021deconditional} to our model, using a globally observed 2D field that would mediate the learning between unmatched AOD and vertically resolved predictors. Such addition would be particularly welcome as it would allow to predict aerosol vertical profiles even at locations where AOD is not observed but vertically resolved predictors are.

In future work we intend to apply our model to AOD arising from satellite observations and meteorological predictors from reanalysis data. Since observations of groundtruth extinction coefficient are not available in 2D satellite products, we intend to collocate observations from MODIS 2D AOD product~\cite{remer2005modis} with CALIOP vertical lidar measurements~\cite{winker2013global} to validate the model.

Finally, a different exciting use-case of the proposed methodology can be considered. If working with climate model data only, then one can choose to use any vertically resolved aerosol tracers as predictors. These include mass and number concentrations which are simulated for several aerosol modes (nucleation, aitken, coarse and accumulation) and species (dust, sea salt, sulfate, black and organic carbon). Leveraging the interpretability of the GP covariance function, the kernel hyperparameters can then indicate which mode/specie did contribute to predictions, hence providing a tool to better understand factors that modulate the optical properties of aerosols.

\newpage

\paragraph{Acknowledgments}
SB, DWP, AN, DS receive funding from the European Union’s Horizon 2020 research and innovation programme under Marie Skłodowska-Curie grant agreement No 860100. DWP acknowledges funding from NERC project NE/S005390/1 (ACRUISE). SS receives funding from the Engineering and Physical Sciences Research Council (EPSRC) via the UK Research and Innovation (UKRI) Centre for Doctoral Training in Application of Artificial Intelligence to the study of Environmental Risks (AI4ER, EP/S022961/1).

\paragraph{Data and Code Availability}
%A statement about how to access data, code and other materials allowing users to understand, verify and replicate findings --- e.g. Replication data and code can be found in Harvard Dataverse: \verb+\url{https://doi.org/link}+.
The data and code that support the findings of this study are openly available at \url{https://github.com/shahineb/aodisaggregation}.

\section*{Nomenclature}

\begin{table}[h!]
    \caption{Acronyms}
    \begin{tabular}{ll}
        ACIs & Aerosols-cloud interactions \\
        AOD & Aerosol optical depth \\ 
        CCN & Cloud condensation nuclei \\
        ELBO & Evidence lower-bound \\
        ICI & Integrated calibration index \\
        GP & Gaussian process \\
        SV & Shapley value \\
    \end{tabular}
\end{table}

\begin{table}[h]
    \caption{Greek letters}
    \begin{tabular}{ll}
        $\alpha$ & Truncated credible interval size \\
        $\gamma_1$ & Spatiotemporal kernel variance hyperparameter \\
        $\gamma_2$ & Meteorological predictors kernel variance hyperparameter \\
        $\eta$ & Generic notation for the mean parameter of a log-normal distribution \\
        $\eta_i$ & Mean parameter of the log-normal observation model for the $i$\textsuperscript{th} column \\
        $\hat\eta_i$ & Approximated mean parameter of the log-normal observation model for the $i$\textsuperscript{th} column \\
        $\lambda$ & Generic notation for wavelength \\
        $\mu$ & Generic notation for the location parameter of a log-normal distribution \\
        $\bmu_\bfw$ & Mean vector of the variational distribution \\
        $\boldsymbol{\bar\mu}_{(\cdot)}$ & Mean vector of the posterior variational distribution evaluated at $(\cdot)$ \\
        $\nu$ & Matérn covariance function order \\
        $\sigma$ & Scale parameter of the AOD log-normal observation model \\
        $\sigma_\text{ext}$ & Scale parameter of the extinction coefficient log-normal observation model \\
        $\bSigma_\bfw$ & Covariance matrix of the variational distribution \\
        $\boldsymbol{\bar \Sigma}_{(\cdot)}$ & Covariance of the posterior variational distribution evaluated at $(\cdot)$ \\
        $\tau$ & Generic notation for the observed AOD \\
        $\tau_i$ & Observed AOD for the $i$\textsuperscript{th} column \\
        $^s\!\tau_i$ & Spatially smoothed AOD observation for the $i$\textsuperscript{th} column \\
        $\btau$ & Vector of observed AOD for the $n$ columns \\
        $\varphi$ & Prior over extinction coefficient profile \\
        $^s\!\varphi$ & Rescaled prior over extinction coefficient profile \\
        $\psi$ & Positive link function warping the GP \\
    \end{tabular}
\end{table}

\begin{table}[h!]
    \caption{Latin letters}
    \begin{tabular}{ll}
        $b_\text{ext}$ & Extinction coefficient \\
        $d$ & Dimensionality of the vertically resolved covariates vector \\
        $\cD$ & Dataset \\
        $f$ & Gaussian process \\
        $\f$ & Generic notation for a realization of the GP at vertically resolved covariates $x|h$ \\
        $\bf$ & Realization of the GP over all vertically resolved inputs in the dataset \\
        $\bf_i$ & Realization of the GP over all vertically resolved inputs from the $i$\textsuperscript{th} column \\
        $h$ & Generic notation for height \\
        $h_i^{(j)}$ & Altitude of the $j$\textsuperscript{th} vertically resolved covariates vector for the $i$\textsuperscript{th} column \\
        $H$ & Generic notation for the total height of an atmospheric column \\
        $\bI_p$ & Identity matrix of size $p$ \\
        $k$ & GP covariance function or kernel \\
        $\bfK_{\bfw\bfw}$ & Covariance matrix evaluated at inducing locations \\
        $\ell_{(\cdot)}$ & Kernel lengthscale hyperparameter for variable $(\cdot)$ \\
        $L$ & Idealized exponential profile lengthscale \\
        lat & Generic notation for latitude \\
        lon & Generic notation for longitude \\
        $m$ & GP mean function \\
        $m_i$ & Number of observed altitude levels for the $i$\textsuperscript{th} column \\
        $M$ & Total number of vertically resolved observations across the dataset \\
        $n$ & Number of columns observed in the dataset \\
        $N(\alpha)$ & Percentage of groundtruth observation falling within the $1-\alpha$ credible interval \\
        $p$ & Number of inducing locations \\
        $P$ & Generic notation for pressure \\
        $q$ & Variational distribution \\
        $S$ & Generic notation for supersaturation \\
        $t$ & Generic notation for time \\
        $T$ & Generic notation for temperature \\
        $\bfu$ & Inducing variables of the variational distribution \\
        $w$ & Weighting function for the idealized exponential profile \\
        $\bfw$ & Inducing locations of the variational distribution \\
        $x$ & Generic notation for the vertically resolved covariates vector \\
        $x_i^{(j)}$ & Observed vertically resolved covariates at altitude $h_i^{(j)}$ for the $i$\textsuperscript{th} column \\
        $\bfx$ & Vector of all vertically resolved covariates from the dataset \\
        $\cX$ & Space in which the vertically resolved covariates vector take values \\
    \end{tabular}
\end{table}

\newpage
\bibliographystyle{unsrtnat}
\bibliography{references}

\newpage
\appendix

\section{Matérn covariance}\label{appendix:matern-covariance}

The Matérn covariances are a class of stationary covariance functions widely used in spatial statistics. The Matérn-$\nu$ covariance between two points $x, x'\in\RR$ is given by
\begin{equation}
    C_\nu(x, x') = \frac{2^{1-\nu}}{\Gamma(\nu)}\left(\sqrt{2\nu}\frac{|x-x'|}{\ell}\right)^{\nu}K_\nu\left(\sqrt{2\nu}\frac{|x-x'|}{\ell}\right),
\end{equation}
where $\Gamma$ is the gamma function, $K_\nu$ is the modified Bessel function and $\ell > 0$ is a lengthscale hyperparameter.

The covariance function expression considerably simplifies for $\nu = p + 1/2$ where $p\in\NN$. For example, for $\nu=1/2$ ($p=0$) and $\nu=3/2$ ($p=1$) we have
\begin{align}
    C_{1/2}(x, x') & = \exp\left(-\frac{|x-x'|}{\ell}\right) \\
    C_{3/2}(x, x') & = \left(1 + \sqrt{3}\frac{|x-x'|}{\ell}\right)\exp\left(-\sqrt{3}\frac{|x-x'|}{\ell}\right)
\end{align}

When $x, x'\in\RR^d$, the distance $|x - x'|$ can be substituted by the norm $\|x - x'\| = \sqrt{\sum_{i=1}^d (x_i - x_i')^2}$. The covariance is called an \emph{automatic relevance determination} (ARD) kernel when each dimension has its own independent lengthscale parameter $\ell_i > 0$. For example, the Matérn-$1/2$ and Matérn-$3/2$ ARD kernel write
\begin{align}
    C_{1/2}(x, x') & = \exp\left(-\sqrt{\sum_{i=1}^d \frac{(x_i - x_i')^2}{\ell_i}}\right) \\
    C_{3/2}(x, x') & = \left(1 + \sqrt{3\sum_{i=1}^d \frac{(x_i - x_i')^2}{\ell_i}}\right)\exp\left(-\sqrt{3\sum_{i=1}^d \frac{(x_i - x_i')^2}{\ell_i}}\right).
\end{align}

When a Matérn-$p + 1/2$ covariance function is used as a kernel for a GP, draws from the GP are $p$ times continuously differentiable (with convention that $0$ times means simply continuous).

\section{Additional sliced plots}\label{appendix:additional-plots}

% comment out when paper ready --- slows down compiling
\begin{figure}[t!]
    \centering
    \includegraphics[width=0.91\textwidth]{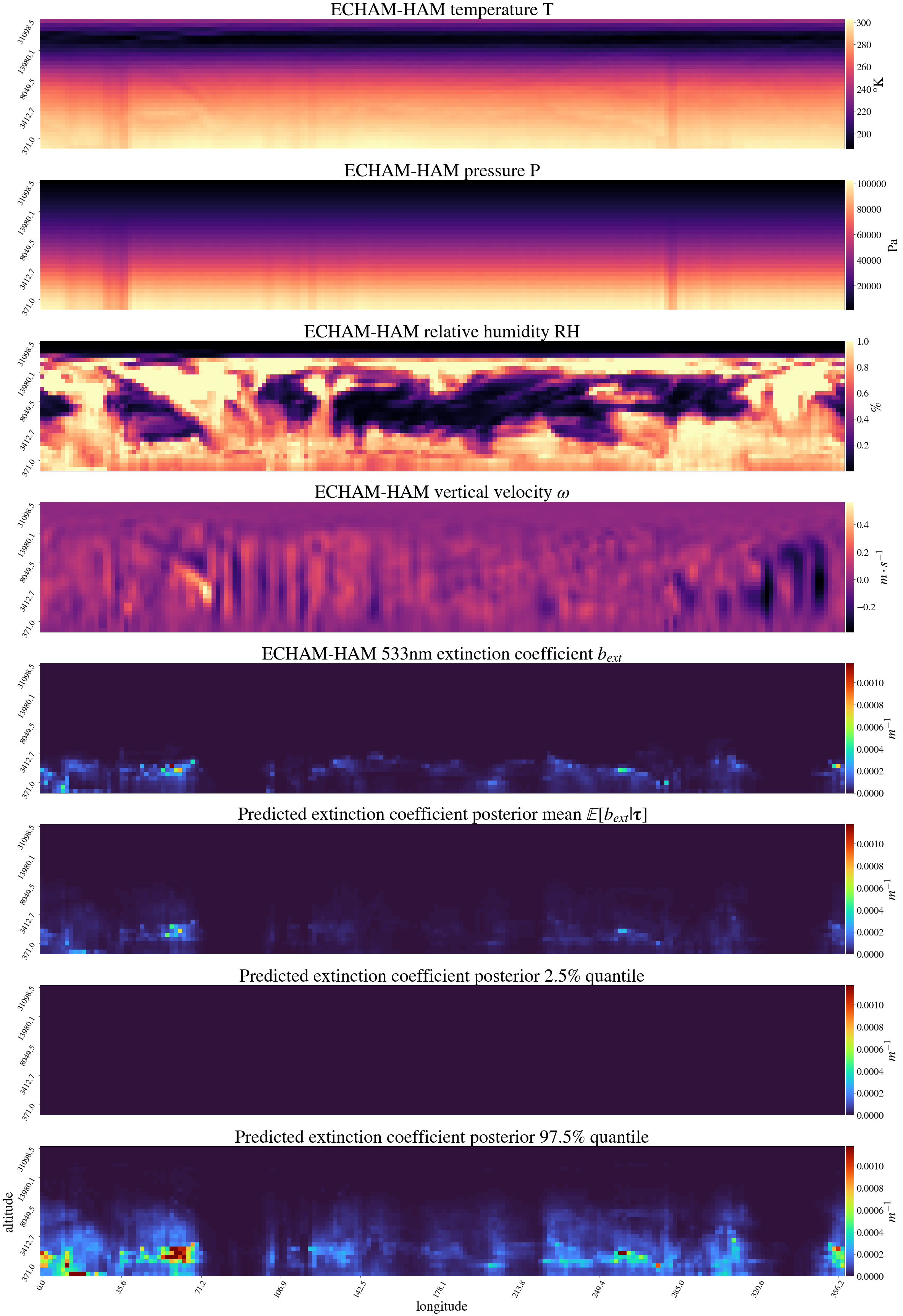}
    \caption{Vertical slices at latitude -0.93$^\circ$ of meteorological predictors ($T, P, \text{RH}, \omega$), groundtruth extinction coefficient, predicted extinction coefficient posterior mean, 2.5\% and 97.5\% quantiles of the predicted extinction coefficient posterior distribution}
    \label{fig:slices-1}
\end{figure}

\begin{figure}[t!]
    \centering
    \includegraphics[width=0.91\textwidth]{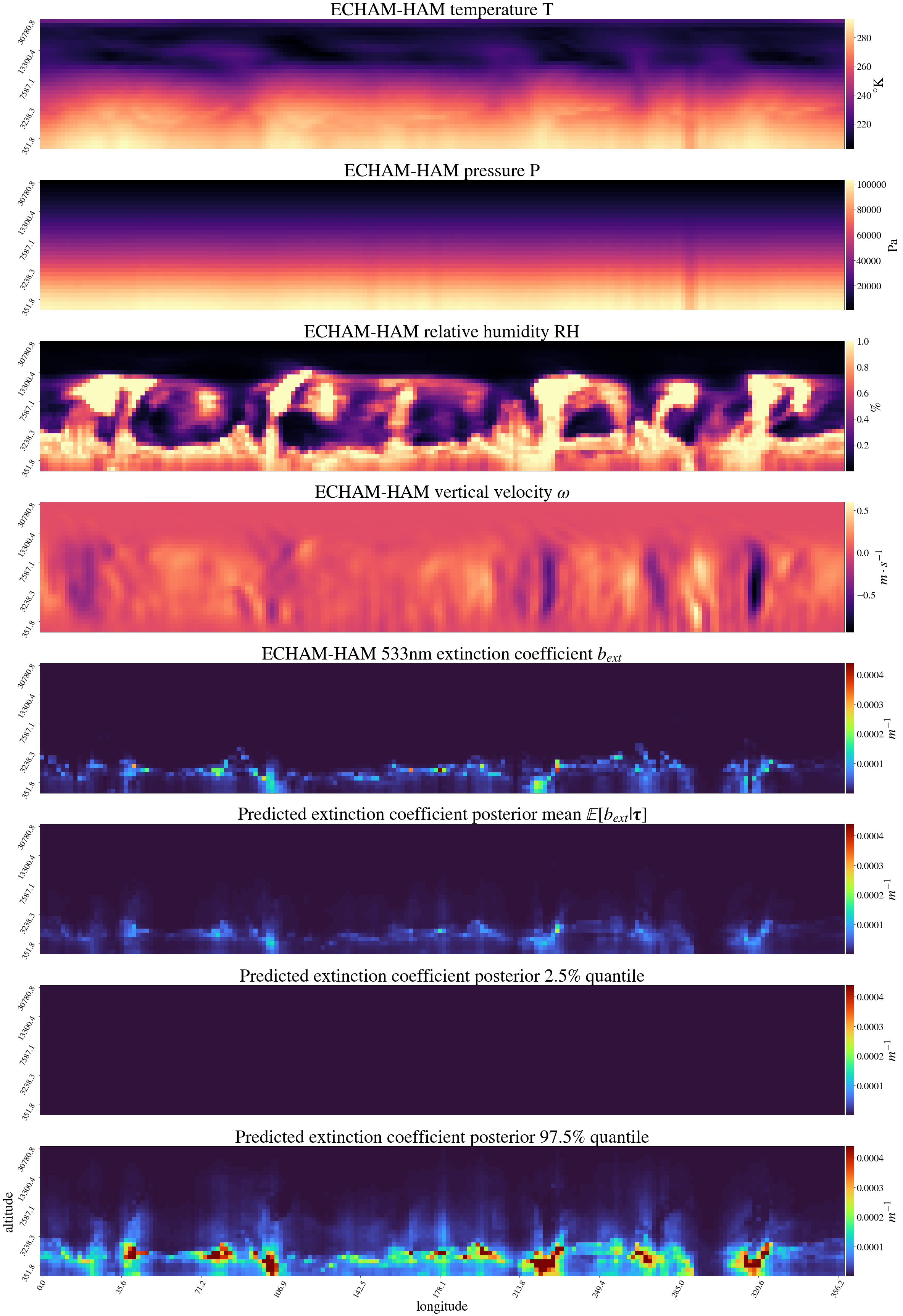}
    \caption{Vertical slices at latitude -38.2$^\circ$ of meteorological predictors ($T, P, \text{RH}, \omega$), groundtruth extinction coefficient, predicted extinction coefficient posterior mean, 2.5\% and 97.5\% quantiles of the predicted extinction coefficient posterior distribution}
    \label{fig:slices-2}
\end{figure}

\end{document}

%% file: lettres_maths.tex
\def\NN{{\mathbb N}}    %naturels
\def\RR{{\mathbb R}}    %r�els
\def\EE{{\mathbb E}}    % espérance
\def\11{{\mathbf 1}}    % indicatrice

        \def\cN{{\mathcal N}}     \def\cO{{\mathcal O}}  \def\cD{{\mathcal D}}            \def\cL{{\mathcal L}}  \def\cX{{\mathcal X}}   

\def\bbf{\mathbf{f}}

\def\bw{\mathbf{w}}

        \def\bI{\mathbf{I}}

\def\EE{\mathbb{E}}

\def\cD{\mathcal{D}}
\def\cN{\mathcal{N}}

\def\cO{\mathcal{O}}

\def\bfx{\mathbf{x}}

\def\bfw{\mathbf{w}}
\def\bfu{\mathbf{u}}

\def\f{\mathrm{f}}

\def\btau{\boldsymbol{\tau}}
\def\bmu{\boldsymbol{\mu}}
\def\bSigma{\boldsymbol{\Sigma}}
\def\beps{\boldsymbol{\epsilon}}
\def\bfK{\mathbf{K}}

\def\d{\,{\mathrm d}}

\definecolor{pink}{HTML}{f77189}
\definecolor{orange}{HTML}{ce9032}
\definecolor{vert}{HTML}{32b166}
\definecolor{turquoise}{HTML}{36ada4}
\definecolor{blue}{HTML}{39a7d0}
\definecolor{olive}{HTML}{97a431}
\definecolor{teal}{cmyk}{0.84,0.13,1,0.33}

\newtcolorbox{CatchyBox}[2][]{
lower separated=false,
colback=white!80!teal,
colframe=white, 
fonttitle=\bfseries,
colbacktitle=teal!40!white,
coltitle=black,
enhanced,
attach boxed title to top left={xshift=.02\linewidth,yshift=-4mm},
title=#2,#1
}